\documentclass[amsmath,amsfonts,amssymb,twocolumn,nofootinbib,floatfix]{revtex4-1}
\usepackage{graphicx}
\usepackage{verbatim}
\usepackage{hyperref}
\usepackage{xcolor}

\begin{document}

\title{\large{On goodness-of-fit tests for arbitrary multivariate models}}
\author{\normalsize{Lolian Shtembari}}
\email{lolian@mpp.mpg.de}
\author{\normalsize{Allen Caldwell}}
\affiliation{Max Planck Institute for Physics, Munich, DE 80805}

\begin{abstract}
\noindent Goodness-of-fit tests are often used in data analysis to test the agreement of a distribution to a set of data. 
These tests can be used to detect an unknown signal against a known background or to set limits on a proposed signal distribution in experiments contaminated by poorly understood backgrounds.
Out-of-the-box non-parametric tests that can target any proposed distribution are only available in the univariate case.
In this paper, we discuss how to build goodness-of-fit tests for arbitrary multivariate distributions or multivariate data generation models.
\end{abstract}

\maketitle

\section*{Introduction}

\noindent Goodness-of-fit tests are often used in data analysis to test the agreement of a distribution to a set of data. 
These tests can be used to detect an unknown signal against a known background or to set limits on a proposed signal distribution in experiments contaminated by poorly understood backgrounds.
Out-of-the-box non-parametric tests that can target any proposed distribution are only available in the univariate case: the Kolmogorov-Smirnov (KS) test \citep{Kolmogorov}, the Anderson-Darling (AD) test \citep{AD} or the Recursive Product of Spacings (RPS) test \citep{RPS}.
In this paper, we discuss how to build goodness-of-fit tests for arbitrary multivariate distributions or multivariate data generation models.
The resulting tests perform an unbinned analysis and do not need any trials factor or look-elsewhere correction since the multivariate data can be analyzed all at once.
The proposed distribution or generative model is used to transform the data to an uncorrelated space where the tests are developed. 
Depending on the complexity of the model, it is possible to perform the transformation analytically or numerically with the help of a Normalizing Flow algorithm.

The flexibility of targeting vastly different univariate distributions is made possible by the probability integral transformation \citep{pearson_1902, pearson_1933}.
We start by reviewing this transformation in the univariate case and then extend it to the multivariate case.
We then discuss different ways of performing a multivariate uniformity test and how to adapt this tool in the case of signal discovery or setting upper limits.

Finally we consider examples for each application in order to test the sensitivity of our methods.

\subsection*{Univariate probability integral transformation}

\noindent Given $m$ univariate samples $\{x_i\}$ assumed to be independent and identically distributed (i.i.d.) according to a known distribution, $f(x)$, we can perform quantitative tests based on the probability integral transformation.
Considering only continuous distributions $f(x)$ with cumulative $F(x)$, we first transform the samples onto the unit interval $[0,1]$ via $u_i = F(x_i)$. 
This reduces the task at hand to test transformed samples $\{u_i\}$ being distributed according to the standard uniform distribution $\mathcal{U}(0,1)$.
Many tests have been developed for this univariate distribution. The take-away message from the univariate case is that, in order to develop a test statistic, it is easier to do so in a standardized space, such as the uniform interval $[0,1]$. 

\section*{Multivariate probability integral transformation}

\noindent 
Much like the univariate case, the goal in multivariate cases (in $n$ dimensions) is to develop uniformity tests in the unit hyper-cube  $[0,1]_n$.
In order to target any given multivariate distribution $\textbf{M}$, we need to transform the probability space described by $\textbf{M}$ into $[0,1]_n$. 
This transformation can be easy or difficult depending on the distribution $\textbf{M}$, specifically depending on the correlation among the dimensions of $\textbf{M}$. 
In the following we show how to perform the transformation into the unit hyper-cube in three main cases: first, distributions comprised of uncorrelated dimensions are considered, moving then to distributions with correlated dimensions or sample generating processes for which a probabilistic model is not available. Finally hierarchical models are discussed.

\subsection*{Independent dimensions}

\noindent If the dimensions of the proposed distribution $\textbf{\textit{M}}$ are all independent of each other, then $\textbf{\textit{M}}$ is just a composition of $n$ independent univariate distributions:

\begin{equation}
    \textbf{\textit{M}} = \left[ M_1, M_2, .., M_n \right]
\end{equation}

\noindent where $M_j$ is the distribution of the $j$-th dimension. 
Much like the univariate case, it is possible to transform the $j$-th component of each sample using the corresponding cumulative distribution function $F_{M_j}$. 
Thus, the transformation of sample $\textbf{\textit{x}}_i = \left( x_{i,1}, x_{i,2}, .., x_{i,n} \right)$ in $[0,1]_n$ is simply:

\begin{equation}
    \textbf{\textit{u}}_{i} = \left[ u_{i,1}, u_{i,2}, .., u_{i,n} \right] = \left[ F_{M_1}(x_{i,1}), .., F_{M_n}(x_{i,n}) \right]
\end{equation}

\subsection*{Correlated dimensions and generative models}

\noindent If the dimensions of the distribution to be compared to the data are not mutually independent, then it might be difficult to write down a transformation to the hyper-cube. 
This is still possible when dealing with nicely behaved distributions, such as a multivariate Normal distribution whose covariance matrix is not diagonal, but that might not be the case for a more complex distribution, such as a weighted sum of distributions. 
In such cases, it is possible to learn the transformation to the unit hyper-cube by using a Normalizing Flow (NF) which can perform a whitening of the distribution; i.e., transform the distribution so that it becomes a diagonal multivariate Normal distribution in the new coordinates. 
Once the original distribution is transformed in this way, it is then possible to further transform it to the unit hyper-cube one component at a time as shown earlier.

\begin{figure}[h]
\centering
\includegraphics[width=0.49\textwidth]{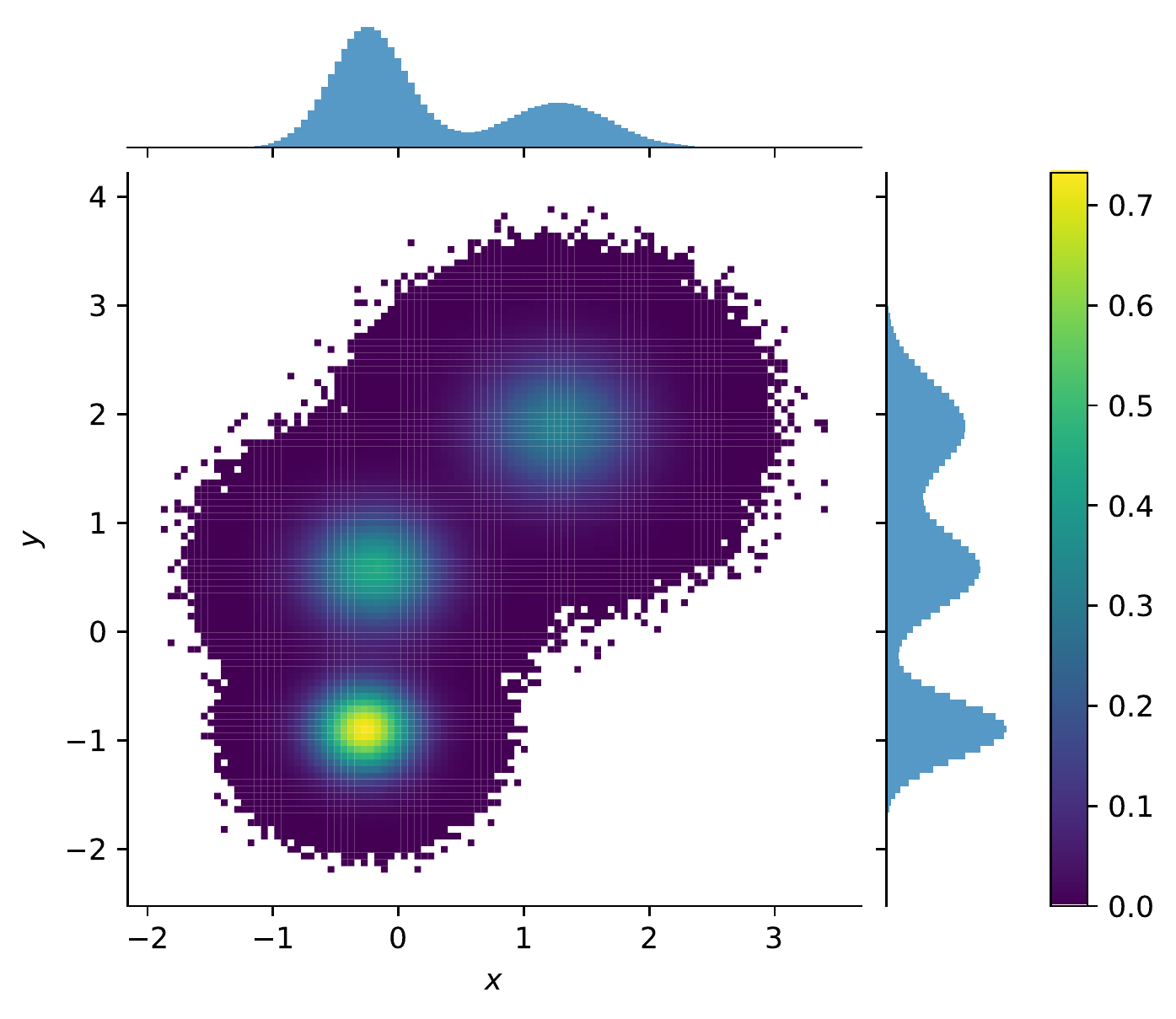}
\caption{Sample distribution of the sum of three two-dimensional Gauss distributions.}
\label{fig:sample_distribution}
\end{figure}

\begin{figure}[h]
\centering
\includegraphics[width=0.49\textwidth]{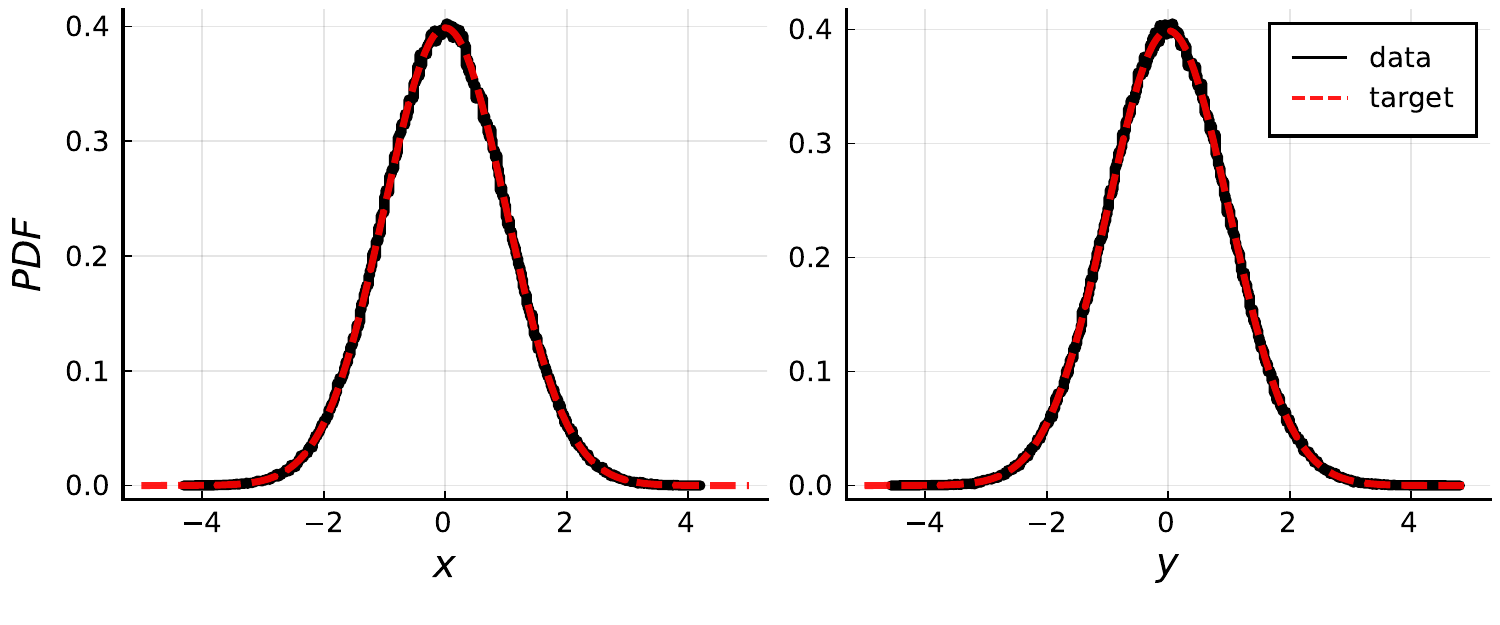}
\caption{Whitened marginal distributions after transforming with the Normalizing Flow.}
\label{fig:normalized_marginals}
\end{figure}

The Normalizing Flow (NF) is made up of a neural network which is trained using samples from the proposed distribution $\textbf{\textit{M}}$. 
The samples needed for training can be obtained from an associated generative model or by sampling $\textbf{\textit{M}}$ using a Markov chain Monte Carlo. 
The use of the generative model is particularly interesting because it allows to train the NF without having a normalized distribution or any model at all. 
In such cases, the NF is learning the associated distribution and the transformation all at once. \citep{9089305, neural_spline_flows} offer a nice review of the theory and some of the many applications of Normalizing Flows. 
In order to show the feasibility of this approach, a proof of principle example is presented where a Normalizing flow is used to whiten data sampled from a sum of three two-dimensional Normal distributions. 
A sampled distribution is depicted in Fig.~\ref{fig:sample_distribution} and the resulting marginal distributions of the whitened samples are shown in Fig.~\ref{fig:normalized_marginals}.
The Normalizing Flow used for this example was adapted from \citep{neural_spline_flows}.

\subsection*{Hierarchical models}

\noindent Given a hierarchical model, the distribution of some components of the data is dependent of the values of other components, which are referred to as hyper-parameters of the model. 
If the hyper-parameters are mutually independent or if a transformation to the unit hyper-cube is available for their distribution and if the same is true for all the dependent parameters at each layer of depth of the hierarchical model, then it is possible to transform the whole distribution into the unit hyper-cube in stages. 

Consider for example a 2 layer hierarchical model producing distributions $\textbf{\textit{M}} = \left[ \textbf{\textit{M}}_1, \textbf{\textit{M}}_2(\textbf{\textit{M}}_1) \right]$. 
$\textbf{\textit{M}}_1$ models the distribution of the hyper-parameters $\textbf{\textit{x}}^{high}$ of the model and these components can be transformed to the corresponding uniform unit hyper-space using the associated function $T_{\textbf{\textit{M}}_1}$. 
The distribution of the dependent parameters $\textbf{\textit{x}}^{low}$ is affected by the observed value of the hyper-parameters $\textbf{\textit{x}}^{high}$:

\begin{equation}
    \textbf{\textit{x}}^{low}_i \sim \textbf{\textit{M}}_2 (\textbf{\textit{x}}^{high}_i)
\end{equation}

\noindent For any given sample $\textbf{\textit{x}}_i$, the value of the hyper-parameters $\textbf{\textit{x}}^{high}_i$ is fixed, so the distribution $\textbf{\textit{M}}_2 (\textbf{\textit{x}}^{high}_i)$ is fully defined and it is possible to compute the corresponding transformation to the unit hyper-space. 
While $T_{\textbf{\textit{M}}_1}$ is sample-independent, $T_{\textbf{\textit{M}}_2}$ is sample-dependent. 
In case of hierarchical models with more layers, this staged transformation approach is to be repeated for each layer.

\section*{Uniformity tests in the unit hyper-cube}

In the following we discuss various methods that allow to perform a multivariate uniformity test by reducing this task to a series of univariate uniformity tests.
These tests are sensitive to non-uniformities in a transformed dataset and their application is twofold: 1) detection of clustering of events against a uniform background, in a discovery scenario, and 2) upper limit on the rate of events corresponding to the uniform component of the data, representative of a proposed signal, against unknown backgrounds.  For the latter, a desired confidence level is set is set in advance.

\subsection*{Projection - Discovery}

\noindent Assume we have $m$ samples within a unit hypercube $\{\textbf{\textit{u}}_i\} \in  [0,1]_n $.  The $n$ components of each sample are assumed independent of one another after the necessary transformations.
The projections of the samples along each axis of the hyper-cube therefore yields $n$ univariate uniformly distributed sets of data: $\{u_{i,j}\}$ for the $j$-th dimension. 
For each one of these projected datasets $\{u_{i,j}\}$ it is possible to perform a uniformity test using a test statistic of choice and condense the information for the $j$-th dimension in one scalar p-value $p_j$. 
Given our assumptions, the expected distribution of each p-value $p_j$ is uniform, and moreover, the p-values will be independent of one another.

On this resulting dataset, $\{p_j\}$, it is possible to perform a uniformity test using a test statistic of choice in order to check whether there are any significant deviations from uniformity. 
The result of this last uniformity check results in one last p-value $p_{final}$ which is the overall p-value of the multivariate goodness-of-fit test.

As pointed out in the discussion above, in order to obtain the intermediate p-values, $\{p_j\}$, and then the final one, $p_{final}$, it is possible to use any test statistic of choice, as long as the chosen statistics preserve the non-correlation among dimensions (results from tests that have a Poisson dependent factor, for example, will be correlated, since the same number of samples is projected on all dimensions). 
What is important is that the distribution of the resulting p-values is uniform. 
This implies that the test statistic used for the evaluation of the intermediate p-values, $\{p_j\}$, does not have to be the same as the one used to evaluate $p_{final}$; as a matter of fact one could also use different tests for different dimensions in the evaluation of $\{p_j\}$, but it might be a more consistent approach to consider all dimensions equally and use the same test for all projections. 

In the previous discussion, we considered a dataset of $m$ samples $\{\textbf{\textit{u}}_i\} \in [0,1]_n $. 
In such a case, if the number of events $m$ is large, it might be appropriate to use a test such as RPS or KS in order to pick up on a signal in any of the projections. 
Afterwards, when considering the $n$ p-values $\{p_j\}$, it could be better to look for outliers, since already one of a few small $p_j$ could be indicative of the presence of a signal in our data. 
In this case, especially when dealing with low-dimensionality spaces ($n$ small), instead of using RPS or KS on the set $\{p_j\}$ it might be more informative to look at the smallest p-value or rather their product in case we want to improve the sensitivity in the presence of multiple small p-values.

\subsubsection*{\textbf{Minimum p-value}}

\noindent As discussed, observing one small p-value might already be enough to point to a possible signal in the data. 
Under the assumption of a uniform distribution of $\{p_j\}$, the distribution of $p_{min} = \min\{p_j\}$ is simply the first Order Statistic, and it follows a Beta distribution \citep{DavidNagaraja:2003}:

\begin{equation}
    p_{min} = \min_{j} \{p_j\} \sim \mathrm{Beta}(1, n)
\end{equation}

\noindent where $n$ is the dimensionality of the original data.
Thus the final p-value is:

\begin{equation}
    p_{final} = F_{\mathrm{Beta}}(p_{min}; 1, n)
\end{equation}

\noindent where $F_{\mathrm{Beta}}(x; a, b)$ is the cumulative distribution function of the Beta distribution with parameters $(a, b)$.

\subsubsection*{\textbf{Product of p-values}}

\noindent Given more than one small p-value $p_j$, looking only at the smallest one might be reductive and we could gain in sensitivity by combining the small p-values together. 
One way of doing so is to consider the product of all p-values:

\begin{equation}
    p_{prod} = \prod_{j=1}^n p_j
\end{equation}

Once again, we expect all $\{p_j\}$ to be uniformly distributed, and the distribution of $p_{prod}$ is known \citep{springer1979algebra}:

\begin{equation}
    P \left( p_{prod} = x; n \right) = \frac{(-1)^{n-1}}{(n-1)!} \left[ \ln(x) \right]^{n-1} 
\label{pdf_prod_n_uni}
\end{equation}

\noindent thus the final p-value $p_{final}$ is:

\begin{equation}
    p_{final} = F \left( p_{prod}; n \right) = p_{prod} \cdot \sum_{j=1}^n \frac{(-1)^{j-1}}{(j-1)!} \left[ \ln(p_{prod}) \right]^{j-1}
\label{cdf_prod_n_uni}
\end{equation}

\subsection*{Projection - Limit setting}

\noindent Several spacings-based tests have been developed for this task in 1 dimension, such as the Maximum-Gap or Optimum-Interval (OI) methods \citep{Yellin_2002}, as well as the Sum-of-Largest-Sorted-Spacings (SLSS) or the Product-of-Complementary-Spacings (PCS) \citep{shtembari2023limit}.

When setting limits in the univariate case, given a test $T$ with cumulative distribution $F_T$, its Poisson-averaged p-value is calculated as:

\begin{equation}
    1 - p = F_{T,Pois}(t_{obs} | \mu) = \sum_{n = 0}^{\infty} F_T(t_{obs} | n) \cdot \frac{\mu^n e^{-\mu}}{n!}
\label{eq:poisson_p_value}
\end{equation}

\noindent where $t_{obs}$ is the observed value of the test-statistic.
Given Eq.~\ref{eq:poisson_p_value} it is possible to find the event rate $\mu_{lim}$ with a confidence level CL such that:

\begin{equation}
    F_{T,Pois}(t_{obs} | \mu_{lim}) = \mathrm{CL}
\end{equation}

\noindent For a more complete discussion regarding how to set upper limits, see reference \citep{shtembari2023limit}.

For the multivariate case, as discussed before, given $m$ uniformly distributed samples $\{\textbf{\textit{u}}_i\} \in [0,1]_n $, we consider the projection of the samples on the $n$ axes, knowing these will be uniformly distributed as well.
For each one of these projected datasets $\{u_{i,j}\}$ it is possible to estimate an upper limit $\mu_j$ on the event rate with confidence level $C_1$.

Out of the upper limits $\{\mu_{j}\}, j=1,..,n$ obtained from each projection, we can use a best of the bunch approach and select the smallest one as the final limit:

\begin{equation}
    \mu_{final} = \min_{j} \{ \mu_j \}
\end{equation}

\noindent At this point we must consider the confidence level $C_n$ associated with this estimate.
If the projected limits $\{\mu_{j}\}$ were completely independent of one another, then we might consider that selecting the smallest limit amounts to a resulting confidence level $C_n$ equal to the product of $n$ Bernoulli variables with rate $C_1$, thus:

\begin{equation}
    C_n = (C_1)^n
\label{eq:naive_relation}
\end{equation}

\noindent Under this assumption, we could easily select the confidence level $C_1$ of the individual projection limit estimations in order to ensure that $C_n$ is equal to the desired value.

This assumption is however not correct. 
Although the distribution of the projected events on each axis is independent, the number of samples projected on each axis is not: if there are $m$ samples in the multi-dimensional space then there will be $m$ samples on each projected dataset $\{u_{i,j}\}, j=1,..,m$.
In order to set a limit we consider both the distribution of events and the total number of events, merging a goodness-of-fit test with a Poisson test.
Since all projected datasets $\{u_{i,j}\}$ share the same number of events, this introduces a correlation in the Poisson statistic part of each limit-setting estimation, rendering all resulting limits correlated.

Although the projection-independence-assumption is not valid if applied after the Poisson-averaging, it is possible to calculate the corrections necessary to ensure the desired final confidence level $C_n$.
We assume that $C_n$ is a function of the projection specific confidence level $C_1$ and that it is dependent on the value of the reconstructed limit $\mu_{final}$, for a given number of dimensions $n$: $C_n(\mu_{final}, C_1 | n)$.
If we seek a specific Confidence Level CL, then we need to find the value of $C_1$ that for the resulting best limit $\mu_{final}$ yields: 

\begin{equation}
    C_n(\mu_{final}(C_1), C_1| n) = \mathrm{CL}
\label{eq:refined_idea}
\end{equation}

This equation is just a one-dimensional root finding problem in $C_1$ which can be solved iteratively (for example using a Bisection algorithm) by estimating the error at $\mu_{final}(C_1)$ for a proposed value of $C_1$.
The estimation of the error rate can be done via Monte-Carlo simulations, producing data according to a uniform distribution in the $n$-dimensional hypercube, since the Eq.~\ref{eq:refined_idea} only needs to hold in this nominal case.

Although this procedure might seem complicated, it is easy to devise and can be performed well before any real analysis has to be run, during the method validation phase, allowing for the tabulation, interpolation and sharing of $C_n(\mu_{final}, C_1 | n)$.
We have calculated the exact correction for the SLSS method and an approximate correction for the OI method up to five dimensions.

\subsection*{Product of Complementary Spacings - Limit setting}

\subsubsection*{Best projection}

\noindent As discussed above, if one calculates the Poisson-averaged p-value on each projected dataset and then chooses the most significant value, a correction needs to be calculated to account for the correlation of these values due to the fixed number of samples on each axis.
In order to avoid this problem, if the definition of the test-statistic chosen allows it, it is possible to perform the selection of the best p-value before averaging with a Poisson distribution.
In such a case it would be trivial to calculate the correct confidence level without having to resort to numerical corrections.

The Product of Complementary Spacings, PCS, is defined as \citep{shtembari2023limit}:

\begin{equation}
    T(\{u_i\}) = - \sum_{i=1}^{n+1} \log(1 - u_{i} + u_{i-1})
\end{equation}

\noindent for a univariate ordered set of data $\{u_i\}$ where $u_0 = 0$ and $u_{n+1} = 1$.
For each of the projected datasets one can compute the corresponding value of the test $T_j$ and its p-value (here $p_j = F_T(T_j)$).
The $n$ projected p-values $p_j$ form an order statistic with uniform distribution. 
If we were to select the largest $F_T(T_j)$, its distribution would be simply:

\begin{equation}
    f \left( \max_j(F_T(T_j)) \right) = \mathrm{Beta}(n, 1).
\end{equation}

\noindent Given the test-statistic values $T_j$ for each projection, the Poisson-averaged p-value of the largest one, $T_{max} = \max_j(T_j)$, is:

\begin{equation}
    F_{T,Pois}(T_{max} | \mu) = \sum_{m = 1}^{\infty} F_{Beta}\left[ F_T(T_{max} | m) \; | n \right] \cdot \frac{\mu^m e^{-\mu}}{m!}.
\end{equation}

\noindent It follows that the upper limit $\mu_{lim}$, with a confidence level CL, is such that:

\begin{equation}
    F_{T,Pois}(T_{max} | \mu_{lim}) = \mathrm{CL}
\end{equation}

\subsubsection*{Sum of projections}

Given the PCS test-statistic values $T_j$ on each projection, instead of selecting the largest, we can consider their sum:

\begin{equation}
    T_{sum} = \sum_{j=1}^{n} T_j
\end{equation}

\noindent which can be interpreted as a product of the product of complementary spacings.
Assuming we know the distribution of  $T_{sum}$ for a fixed number of events $m$, $F(T_{sum}|m)$, then we can compute the Poisson-averaged p-value of this test for a given event rate $\mu$:

\begin{equation}
    F_{Pois}(T_{sum} | \mu) = \sum_{m=1}^{\infty} F(T_{sum} | m) \cdot \frac{\mu^m e^{-\mu}}{m!}
\end{equation}

\noindent Given this definition, it is possible to invert the formula and find the upper limit on the event rate up to a desired confidence level.
For example, the 90\% confidence level upper limit $\mu_{lim}$ is such that:

\begin{equation}
    F_{Pois}(T_{sum} | \mu_{lim}) = 0.9
\end{equation}

\noindent If $F(T_j | m)$ is known, it is rather easy to compute $F(T_{sum}|m)$.
Since $T_j$ are all i.i.d., the distribution of $T_{sum}$ is just $f_{T, m}$ convolved $n-1$ times with itself:

\begin{equation}
    f(T_{sum} | m) = \underbrace{ f(T | m) \ast f(T | m) \ast .. \ast f(T | m)}_\text{$n$ times}
\end{equation}

\noindent Since $F(T| m)$ has been tabulated in the Julia package SpacingStatistics.jl \citep{SpacingStatistics.jl} and is available as a monotonic cubic spline polynomial function, it is possible to easily obtain its derivative $f(T| m)$, transform it to the Fourier space using an FFT, raise it to the power of $n$ and transform back to the real space to obtain $f(T_{sum} | m)$:

\begin{equation}
    f(T_{sum} | m) = FFT^{-1} \left\{ \left[ 
FFT(f(T | m)) \right]^n \right\}
\end{equation}

\noindent This procedure is used for the tabulated $F_{PCS, m}$ ($m \leq 10^4$). For values of $m$ larger than $10^4$ we use the asymptotic distribution of $F_{PCS, m}$, which is a Gaussian distribution, thus rendering the calculation of the convolution much easier.

These two approaches show how to adapt the PCS test to a multivariate limit-setting scenario, similarly to how the minimum p-value and product of p-values were used in the multivariate discovery case.
Although we discussed the PCS test specifically, these correction apply in general to any test-statistic $T$ calculated where the Poisson-averaging can be calculated as a final step.

\subsection*{Volume transformation method}

\noindent Finally, we consider a different dimensionality reduction strategy.
Given $m$ samples $\{\textbf{\textit{u}}_i\} \in [0,1]_n $, instead of projecting them onto the axes and obtaining $n$ independent sets of univariate data, we can use a dimension-reducing transformation to map them all at once onto a single univariate dataset. To achieve this,
 we calculate the volume contained in the hyper-rectangle defined by its projections simply by taking the product of its coordinates:

\begin{equation}
    v_i = V(\textbf{\textit{u}}_i) = \prod_{j=1}^{n} u_{i, j} \; .
\label{eq:volume_transformation}
\end{equation}

\noindent Calculating the volume in this way for each multivariate sample we obtain a simple univariate dataset: $\{\textbf{\textit{u}}_i\} \xrightarrow[]{V} \{ v_i \}$.
Since the $\{\textbf{\textit{u}}_i\}$ were i.i.d. samples, so are the $\{ v_i \}$ (although not uniformly distributed).
Since $v_i$ is the product of $n$ independent uniform variables, whose distribution is given by Eq.~\ref{pdf_prod_n_uni}, its probability distribution is known.
Using the probability integral transformation, Eq.~\ref{cdf_prod_n_uni}, we can therefore transform $\{ v_i \}$ into a set of uniform i.i.d. samples $\{ z_i \}$.
We can use these to then perform a univariate uniformity test using a test statistic of choice; standard discovery and limit-setting tests can then be used in order to analyse the data.

\section*{Example - $\textbf{\textit{n}}$D Discovery}

\newcommand{\ns}{\ensuremath{\langle n_s \rangle}}
\newcommand{\nb}{\ensuremath{\langle n_b \rangle}}

\noindent Here we illustrate how the proposed goodness-of-fit tests can be used in a scenario where a possible `new physics' model is searched for but it is not wished to specify how the new physics might populate the data space.  It is then to be tested whether the data follows a known distribution, which is a `background' to a possible new signal. After having collected some data, one wants to quantify the goodness-of-fit of the background only distribution to the data and a resulting low p-value could indicate the presence of events distributed according to an additional, previously unknown, signal distribution. 

\subsection*{Multivariate Gaussian signal}

In this example the background is modelled by a simple uniform distribution in the 5-dimensional hyper-cube $[0,1]_5$ and in order to illustrate how the presence of an actual signal (alternative hypothesis) would affect the outcome, additional events are injected, following a multivariate Normal distribution randomly positioned within the hyper-cube with isotropic variance of either $0.01$ or $0.1$.
The number of events is Poisson fluctuated for both background and signal populations, with expected values of $\nb=10^4$ and expected values of $\ns$ ranging up to $10^3$.

\begin{figure}[h]
\centering
\includegraphics[width=0.49\textwidth]{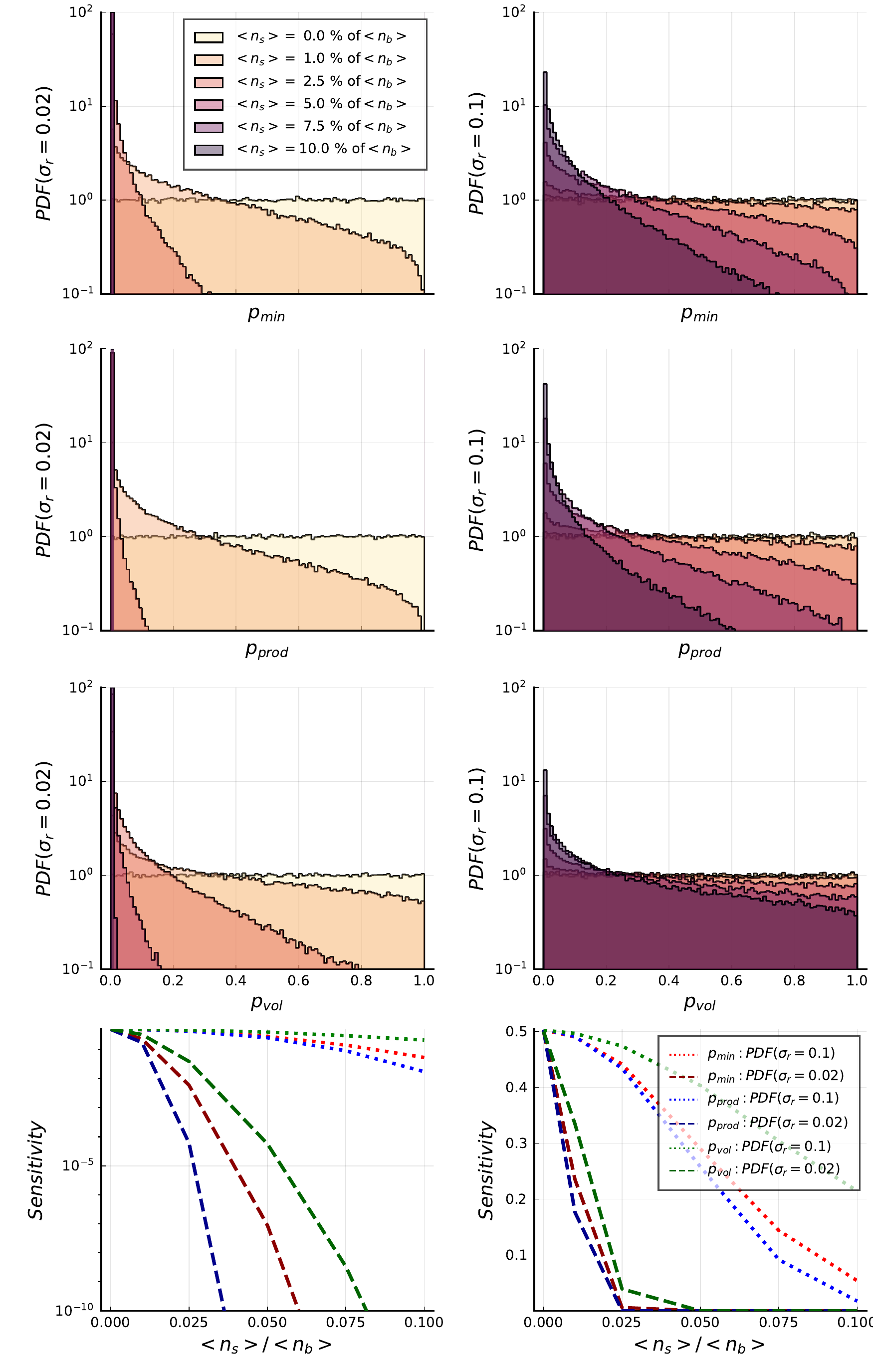}
\caption{Distributions of p-values for background only samples ($\ns=0$) and background plus randomised signal injections from a 5D Gaussian distribution: `narrow' signal with random $\mu \in [0.2, 0.8]$, $\Sigma=0.01 \cdot \mathrm{I}_5$ (left) and `wide' signal with random $\mu \in [0.2, 0.8]$, $\Sigma=0.1 \cdot \mathrm{I}_5$ (right) of varying strength; comparison to the background model for either the minimum p-value statistic (first row), the product of p-values statistic (second row) or the volume-transformed p-value (third row); median p-value (sensitivity) both in linear and logarithmic scale (fourth row).}
\label{fig:sensitivity_example_C}
\end{figure}

The p-value distributions under the assumption of $H^0$ (i.e. only background is present) are shown in Fig.~\ref{fig:sensitivity_example_C}: the results corresponding to the narrow signal ($\Sigma=I_5 \cdot 0.01$) are on the left (first column) and those corresponding to the broad signal ($\Sigma=I_5 \cdot 0.1$) are on the right (second column); the first two rows present p-value distributions calculated using projection methods while the third row presents p-value distributions obtained with the volume transformation method; the fourth row presents the sensitivity of each scenario quantified as the median p-value for each distribution.
Regarding the results of the projection method, the evaluation of the intermediate p-values was performed using the KS test, given the large count rates, while the evaluation of the final p-value, since there are only 5 dimensions, was performed using the two tests previously described, namely the minimum and the product of intermediate p-values, corresponding to the first and second row respectively.
Similarly, the KS test statistic was used in the final uniformity test after performing the volume transformation.

Distributions with no signal ($\ns= 0$) show a flat p-value distribution, as expected, while the distributions of trials with injected signals are trending towards smaller p-values, indicating the worsened goodness-of-fit for the background only model. 
The distributions of trials where the signal has smaller variance (left) are much more skewed towards small p-value compared to those where a larger variance signal was injected (right). 
This shows how the sensitivity of the tests varies when targeting clusters of varying width and strength relative to the background.

In this example, since the signal can be spotted in the projection of multiple dimensions, the product of p-values test (second row) offers the largest rejection probability of the null hypothesis compared to the volume-transformed p-value (third row) or the minimum p-value test (first row).

\subsection*{Multivariate Gaussian-shell signal}

\begin{figure}[h]
\centering
\includegraphics[width=0.49\textwidth]{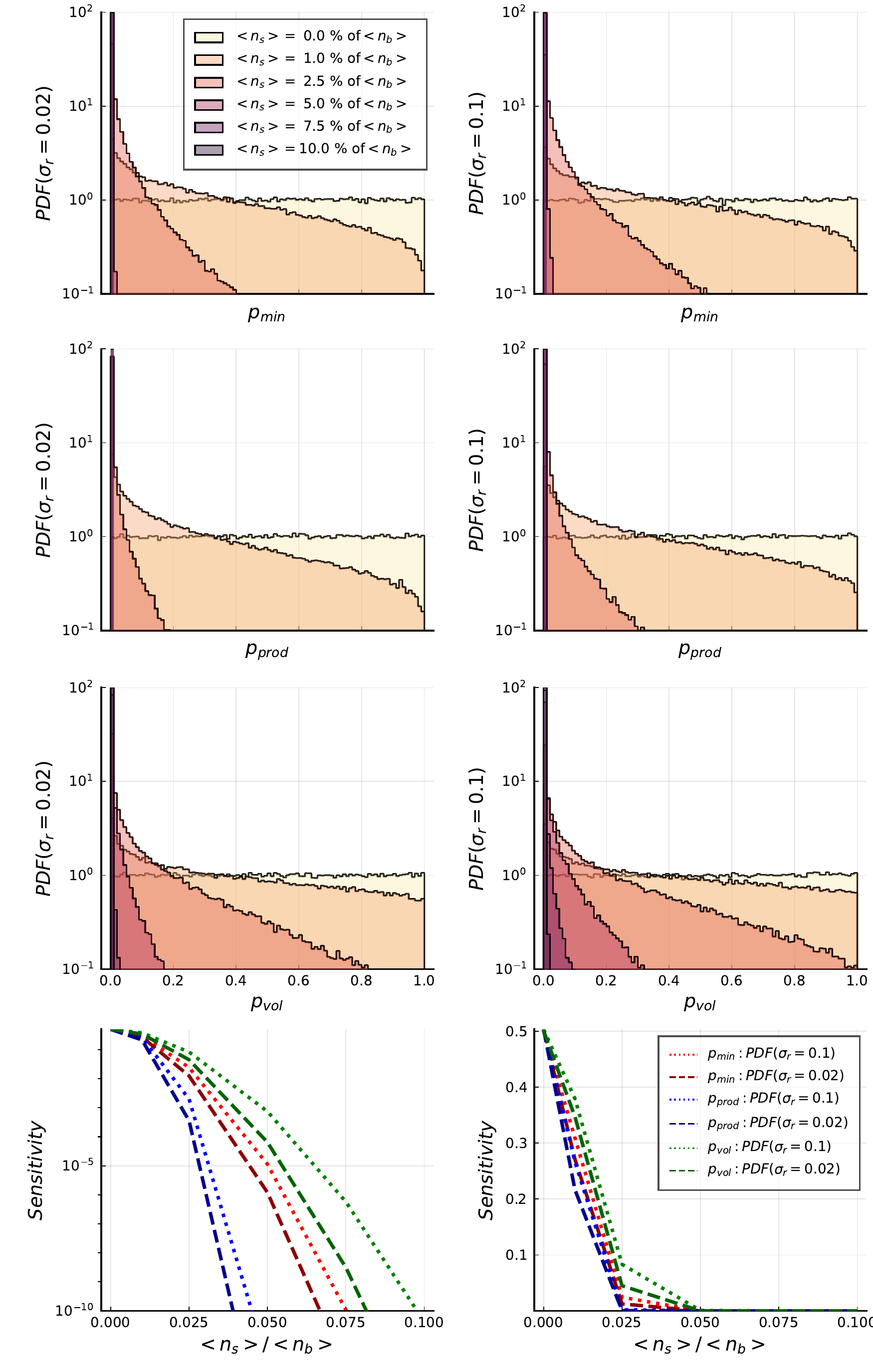}
\caption{Distributions of p-values for background only samples ($\ns=0$) and background plus randomised signal injections from a 5D Gaussian-shell distribution: `narrow' signal with random $\mu \in [0.25, 0.75]$, $r=0.25$, $\sigma_r=0.02$ (left) and "wide" signal with random $\mu \in [0.25, 0.75]$, $r=0.25$, $\sigma_r=0.1$ (right) of varying strength; comparison to the background model for either the minimum p-value statistic (first row), the product of p-values statistic (second row) or the volume-transformed p-value (third row); median p-value (sensitivity) distribution both in linear and logarithmic scale (fourth row).}
\label{fig:sensitivity_example_E}
\end{figure}

Instead of injecting a clustered signal, we assess the sensitivity of our methods in the case of a Gaussian-shell signal.
Our signal is five-dimensional and characterized by a radius $r=0.25$, a radial standard deviation of either $\sigma_r=0.02$ or $\sigma_r=0.1$ and the center of the distribution $\mu$ chosen at random within the hypercube $[0.25, 0.75]_5$.
The results are shown in Fig.~\ref{fig:sensitivity_example_E}.
In this case, we notice that the sensitivity to either signal thickness, $\sigma_r$, is very similar, which shows that all methods are mostly sensitive to the shell-like structure and its radial extension.
Of the three tested methods, the product of p-values shows the highest sensitivity, followed by the minimum p-value and then the volume transformed p-value.

Note that the data in the previous examples were analyzed all in one pass for each trial, meaning that the extracted p-values do not need any corrections for a `trials effect' or `look-elsewhere effect'.
Of course, if one analyzes many separate sets of data, the resulting p-value will need to be corrected as is usually done in the univariate case.

\section*{Example - $\textbf{\textit{n}}$D Limit setting}

\noindent The performance of our proposed methods for limit-setting is explored in a series of
simulated experiments for multivariate sample distributions.
We consider the case where a background model is not present, and only a distribution of counts according to a signal model is available.  In this case, the task is to set a limit on the signal strength of the signal model.

\subsection*{Background-free experiment}

\begin{figure}[h]
\centering
\includegraphics[width=0.49\textwidth]{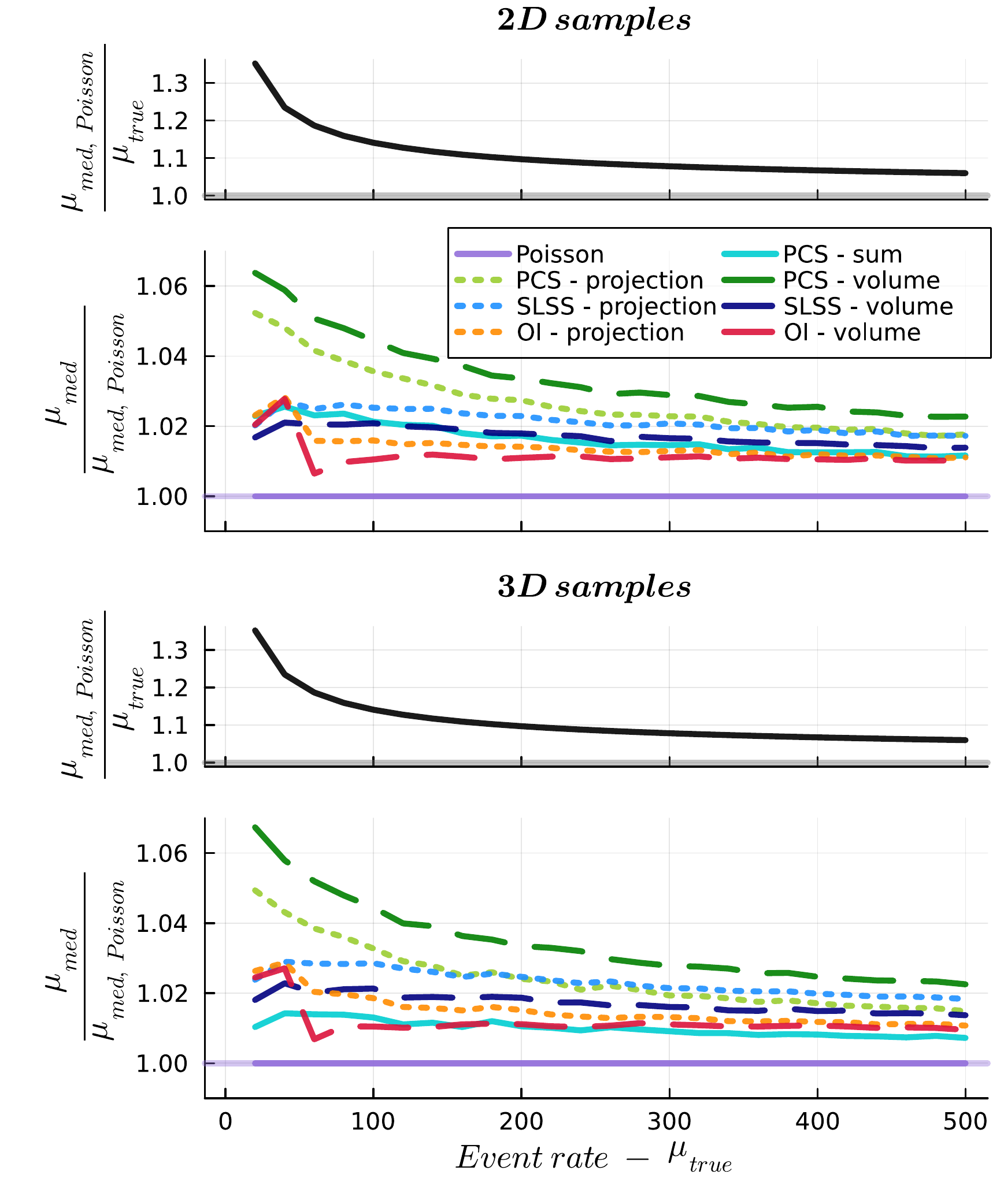}
\caption{Median $CL=0.90$ upper limit for the Poisson test, upper panels, and for tests discussed in the text normalized to the limit from a standard Poisson probability test, lower panels, for 2D (top) and 3D (bottom) uniform signal distributions and no background.}
\label{fig:median_all_ratio_poisson_UNI_SIG}
\end{figure}

\noindent We start by considering the case in which no background contaminates the experiment, in order to estimate the baseline of the different methods. 
Fig~\ref{fig:median_all_ratio_poisson_UNI_SIG} shows the median of the  $CL=0.90$ upper limits on the event rate normalized to the median limit of the Poisson test.
We notice that in this baseline scenario the Poisson test is the best of the bunch, as expected, but it does not drastically outperform the others.


\subsection*{Background-only experiment}

\noindent Next we investigate the case in which a background is present in our simulations and the signal strength is negligible in comparison: this mimics a rare process search in which the signal is absent.

\subsubsection*{\textbf{Exponential distribution}}

\begin{figure}[h]
\centering
\includegraphics[width=0.49\textwidth]{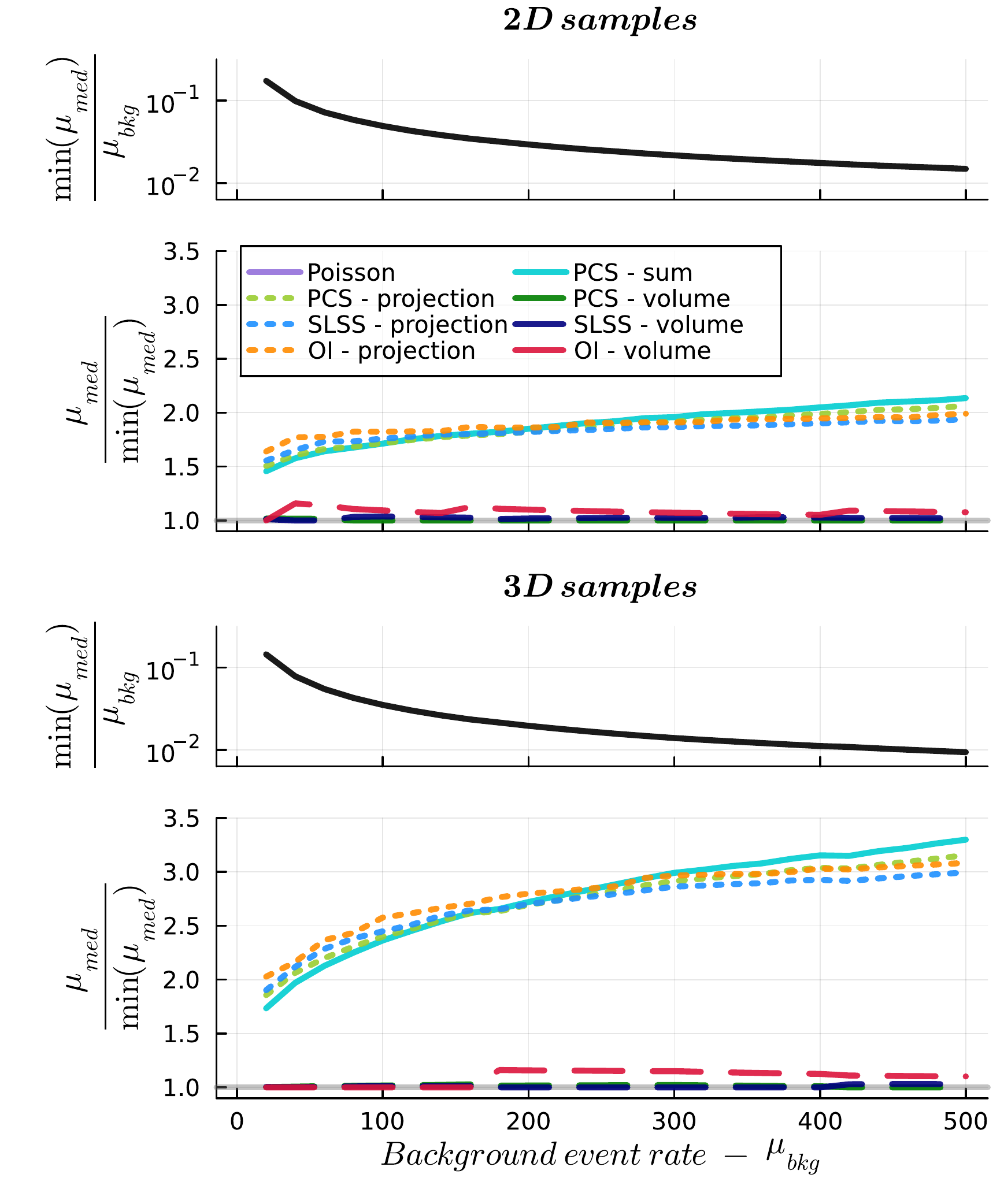}
\caption{Median $CL=0.90$ signal upper limit for the best available test, normalized to the background strength, upper panels, and for tests discussed in the text normalized to the limit from the best test, lower panels, for 2D (top) and 3D (bottom) distributions containing only an exponentially distributed background.}
\label{fig:median_all_ratio_best_EXP_BKG}
\end{figure}

We first consider a background resulting from the product of $n$ independent Exponential distributions of rate $0.1$ in each dimension.

Fig.~\ref{fig:median_all_ratio_best_EXP_BKG} reports the median $CL=0.90$ upper limits of the measured event rate normalized to the smallest median result for a specific background event rate $\mu_{bkg}$. 
Analysing these results, we notice that the volume transformation method provides the best limits, regardless of the test used.
All other projection-based methods perform similarly: in the two-dimensional scenario, the limits are a factor $1.5-2$ worse than the volume transformation results, and in the case of a  three-dimensional distribution a factor $2-3$ worse.

\subsubsection*{\textbf{Gaussian distribution}}

\begin{figure}[h]
\centering
\includegraphics[width=0.49\textwidth]{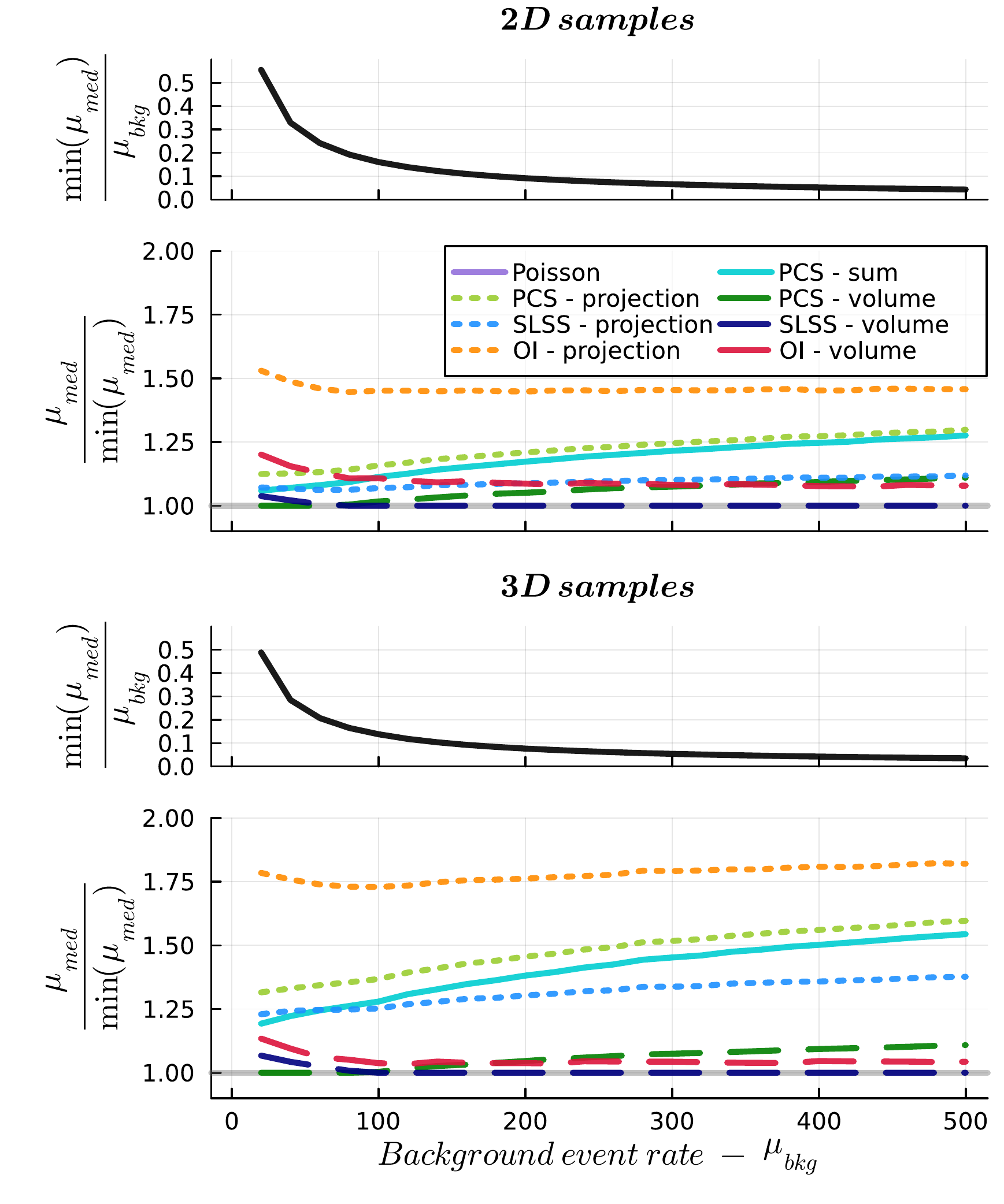}
\caption{$CL=0.90$  upper limit normalized to minimum median result respectively for 2D (top) and 3D (bottom) multivariate Normal distributions with $\Sigma = I \cdot 0.01$ centered in the middle of the hypercube. The upper panels in each case show the best limit result normalized to the background expectation.}
\label{fig:median_all_ratio_best_GAUSS_BKG}
\end{figure}

Next we consider a background distributed according to a multivariate Gaussian centered at the middle of the hypercube and with covariance matrix $\Sigma = I \cdot 0.01$.

Fig.~\ref{fig:median_all_ratio_best_GAUSS_BKG} reports the median $CL=0.90$  upper limits of the measured event rate normalized to the smallest median result for a specific background event rate $\mu_{bkg}$. 
Once again, the volume transformation method provides the best limits, regardless of the test used.
Out of these, the SLSS test is the best of the bunch, since it is better suited to analyse data sets that present multiple disconnected low density regions.

The projection-based methods provide weaker limits: the SLSS and PCS version being up to a factor $1.25 (1.5)$ larger in the 2D (3D) case respectively; the OI test limits are weaker by a factor $1.5 (1.75)$ in the 2D (3D) case respectively.  This is understandable since this test relies only on one low density region to estimate its limit.

\subsubsection*{\textbf{Concave distribution}}
\begin{figure}[h]
\centering
\includegraphics[width=0.49\textwidth]{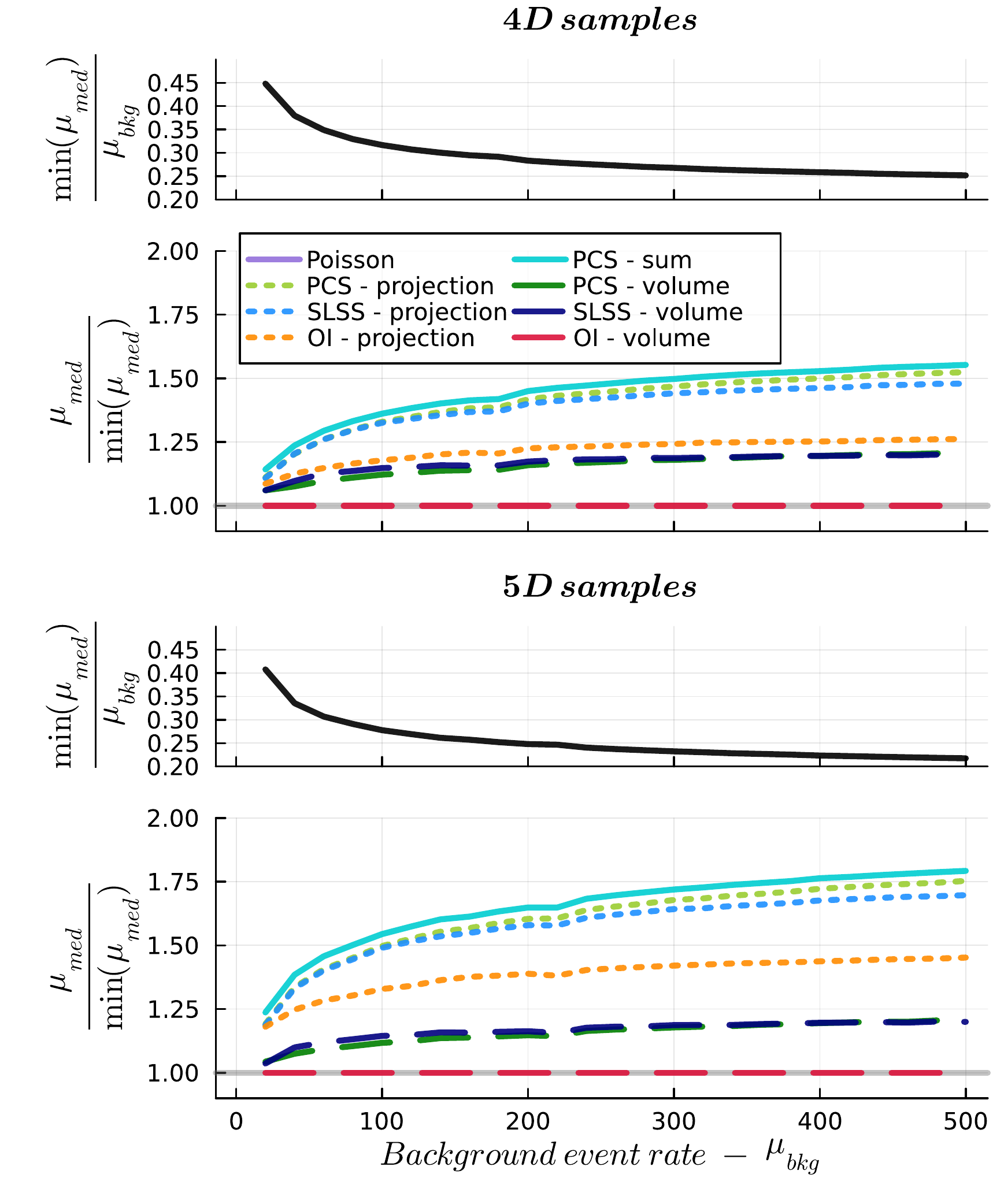}
\caption{$CL=0.90$  upper limit normalized to minimum median result respectively for 2D (top) and 3D (bottom) multivariate Normal distributions for the concave background model. The upper panels in each case show the best limit result normalized to the background expectation.}
\label{fig:median_all_ratio_best_GAUSS_INV_BKG}
\end{figure}

Finally, we consider a bowl shaped background, obtained by reversing the roles of signal and background distribution of the previous example: assuming a uniform background and a Gaussian signal in the real space (truncated to the unit interval $[0,1]$ with $\mu=0.5$ and $\sigma=0.1$), we perform the probability integral transformation with respect to the latter, obtaining a bowl shaped background distribution in the cumulative space.

Fig.~\ref{fig:median_all_ratio_best_GAUSS_INV_BKG} reports the median $CL=0.90$  upper limits of the measured event rate normalized to the smallest median result for a specific background event rate $\mu_{bkg}$. 
In this case we show results for four and five dimensional distributions of events.
We notice that the best results in this case are set by the OI-test with volume transformation.
This is reasonable since there is only one fully connected region of low event-density, namely the basin of the bowl, thus being the best-suited case for the OI test.
The next best results are obtained by the SLSS and PCS volume transformations, which yield no more than 25\% larger limits.
Finally, the projection-based methods yield the most conservative limits, with the OI test being the best of this subset.

\section*{Conclusions}

\noindent We have provided novel non-parametric statistics to perform goodness-of-fit tests targeting any given multivariate distribution or multivariate generative model by means of a transformation to the uniform unit hyper-cube. 
Our approaches allow for unbiased tests, either by considering the volumes identified by each sample or by taking into account their projections.
The tests developed with these methods perform an unbinned analysis of the data and do not need any trials factor or look-elsewhere correction since the multivariate data is analyzed all at once.
These novel methods allow to test for the presence of a signal beyond the known background expectation, or to set a limit on a signal's event rate in cases where the background is not well modeled.
The sensitivity of our proposals were tested in the contest of a mock signal searches.  We have also compared the limit setting capabilities of our methods in simulated rare process searches under a variety of background behaviors.

The test statistics described in this paper are simple to use and code is available to interested users.

\section*{Acknowledgements}

\noindent We thank Dr. Oliver Schulz for helpful discussions and comments. We thank Dr. Oliver Schulz, Dr. Vasyl Hafych and Michael Dudkowiak for their help in the implementation of the Normalizing Flow used in this paper.

\bibliography{main}
\bibliographystyle{PhysRevStyle.bst}

\end{document}